\documentclass[%
 reprint,
 amsmath,amssymb,
 aps,
]{revtex4-1}

\usepackage{graphicx}
\usepackage{dcolumn}
\usepackage{bm}

\begin{document}

\preprint{APS/123-QED}

\title{Extension of the Lieb-Schupp theorem\\to the Heisenberg models with higher order interactions}

\author{Kengo Tanaka$^{1,2}$}
 \altaffiliation[]{}
 \email{tanaka@koran.ac.jp}
\affiliation{%
$^{1}$Information Technology Center, Koran Women's Junior College, Fukuoka, 811-1311, Japan\\
$^{2}$Department of Comprehensive Studies for Life Planning, Koran Women's Junior College, Fukuoka, 811-1311, Japan
}%




\date{\today}

\begin{abstract}
We extend the Lieb-Schupp theorem to the Heisenberg models with higher order interactions 
on non-frustrated or frustrated finite lattices.
These lattices are constructed by even numbered rings with or without crossing bonds 
and have reflection symmetry. 
The results show that all ground states have total spin zero 
in wide interaction parameters region which is not covered with the results of 
the Marshall-Lieb-Mattis type arguments. 
\begin{description}
\item[PACS numbers]
May be entered using the \verb+\pacs{#1}+ command.
\end{description}
\end{abstract}

\pacs{Valid PACS appear here}
\maketitle

\section{\label{sec1}Introduction}
The Heisenberg models with higher order interactions have been discussed from various points of view. 
Hamiltonian of the simplest model consists of bilinear and biquadratic exchange interaction terms. 
The $S=1$ bilinear-biquadratic exchange interaction model has been investigated in
the context of the Haldane conjecture\cite{aklt} and as models of 
one dimensional spin-Peierls material Li$_{2}$VGe$_{2}$O$_{6}$,\cite{millet,mila}  
two dimensional gapless spin liquid material NiGa$_{2}$S$_{4}$,\cite{tsunetsugu,QT1,QT2} 
magnetism in $S=1$ bosons\cite{Yip,imambekov} and three-flavor fermions\cite{flavor} trapped in optical lattices, 
magnetism in iron pnictide superconductors\cite{pnic1,pnic2}, 
and deconfined criticality and Landau forbidden phase transition\cite{HKT,TS,nishiyama}. 
In the case of $S=3/2$ this model is also investigated as a model of the chromium spinel oxides 
$A{\rm Cr}_{2}{\rm O}_{4} (A={\rm Hg}, {\rm Cd}, {\rm Zn})$.\cite{spinel1,spinel2} 
The $S=3/2$ bilinear-biquadratic-bicubic exchange interaction model 
is studied as a model of magnetism in $S=3/2$ fermions trapped in optical lattices \cite{tu,eckert,FKKI} and 
as a resource of the measurement based quantum computer.\cite{wei,miyake} 
In particular the $S=1$ model was extensively studied by theoretical works, 
but models with $S\ge3/2$ and/or bicubic and more higher order interactions are less studied 
and is of importance to the understanding of magnetic properties of the chromium spinel oxides 
and cold atomic gases in optical lattices. 

The Marshall-Lieb-Mattis theorem is one of the most famous exact results of 
quantum spin systems. 
In the case of the antiferromagnetic Heisenberg model on bipartite lattices 
with the same number of sublattice points, 
it proved that ordering energy levels, i.e., the lowest energy level for allowed total spin is monotonically increasing function of total spin and ground state is unique spin singlet.\cite{LM} 
This theorem was extended to the case of the spin-$S$ bilinear-biquadratic exchange interaction model.\cite{munro,parkinson,tanaka} 

The Marshall-Lieb-Mattis theorem has made a lasting contribution to check validity of a huge number of results 
for the numerical studies of quantum spin systems on bipartite lattices by using 
the exact diagonalization, density matrix renormalization group, 
quantum Monte Carlo simulation, etc.,  
and now it serves as guidelines for `{\it real quantum simulators}' envisioned by Richard Feynman.\cite{RQS1,RQS2} 
But this theorem is not applicable to the models on non-bipartite or frustrated lattices. 
In 1999 Lieb-Schupp succeeded to prove that ground states of 
the antiferromagnetic Heisenberg model on checkerboard type of the square lattice with crossing bonds have total spin zero.\cite{lieb-schupp1,lieb-schupp2,schupp} 
Their method use reflection symmetry of Hamiltonian, 
on the other hand, the Marshall-Lieb-Mattis theorem is based on the Perron-Frobenius theorem 
and works well if it can be find suitable unitary transformation 
which leads to same sign of off-diagonal matrix elements of irreducible unitarily transformed Hamiltonian 
satisfying the Perron-Frobenius theorem. 
But it seems that there is no systematic method available to find it so far.
The Lieb-Schupp theorem can be applied to a class of frustrated spin systems on
reflection symmetric lattices, 
but it can not give any information for the degeneracy of the ground state. 

Our purpose in the present paper is an extension of the Lieb-Schupp theorem 
to the Heisenberg models with higher order interactions on finite size lattices 
which are constructed by even numbered rings with or without crossing bonds 
and have reflection symmetry.  
As explained above, 
the Marshall-Lieb-Mattis type argument does not work for non-bipartite lattices. 
Adding antiferromagnetic crossing bonds to even numbered rings 
induces a frustration of N\'{e}el order and breaks bipartiteness of lattices, 
but their reflection symmetry are preserved. 
By using this nature of lattices we will prove that all ground states of these models 
possess total spin zero in wide parameter region 
which is not covered with results of  the Marshall-Lieb-Mattis type arguments. 

This paper is organized as follows. 
In section \ref{sec2}, we introduce some definitions and notation used throughout this paper. 
In section \ref{sec3}, to keep the paper self-contained, 
we explain a basic setup and ideas of the Lieb-Schupp theorem and  
apply this theorem to the models on even numbered rings to prove that 
all ground states have total spin zero. 
In section \ref{sec4}, with Hamiltonian on even numbered rings discussed in section \ref{sec3} as a local Hamiltonian,
we construct global Hamiltonian on two dimensional lattices. 
In particular we treat square and honeycomb lattices with crossing bonds. 
In section \ref{sec5}, we summarize and discuss the results of sections \ref{sec3} and \ref{sec4} 
and comment on the effects of the crossing bonds 
in infinite system of the $S=1$ bilinear-biquadratic exchange interaction model and 
physical realization of ferroquadrupole (spin nematic) phase in magnetic materials. 
\section{\label{sec2}Definition and Notation}
In this paper we study the isotropic spin-$S$ Heisenberg model with up to the $n$-th order $(1\le n\ll\infty)$ interaction term:
\begin{eqnarray}
H_{n}=-\sum_{x\ne y\in\Lambda}\sum_{k=1}^{n}J_{k}(|x-y|)\left(\mbox{\boldmath $S$}(x)\cdot\mbox{\boldmath $S$}(y)\right)^{k}, \label{H}
\end{eqnarray}
on lattice $\Lambda$,  
where $-J_{k}(|x-y|)$ are the $k$-th order interaction coefficients between sites $x$ and $y\in\Lambda$. The summation over $x\ne y\in\Lambda$ counts every pair (once and once only).  
$\mbox{\boldmath $S$}(x)=(S_{1}(x),S_{2}(x),S_{3}(x))$ denotes spin-$S$ operator on site $x$ and
satisfies the usual commutation relations:
\begin{eqnarray}
\left[S_{i}(x),S_{j}(y)\right]={\rm i}\epsilon_{ijk}S_{k}(x)\delta_{xy}.
\end{eqnarray} 
Here we use a usual basis in which $S_{3}(x)$ is diagonalized, 
$S_{1}(x)$ and $S_{3}(x)$ have real matrix elements and $S_{2}(x)$ pure imaginary.

This Hamiltonian can be written as the spin-$S$ isotropic Hamiltonian with up to $2^{n}$-pole interaction term:
\begin{eqnarray}
{\cal H}_{n}=-\sum_{x\ne y\in\Lambda}\sum_{k=1}^{n}I_{k}(|x-y|)\sum_{q=-k}^{k}O_{k,q}(x)O_{k,q}^{\dagger}(y),\label{calH}
\end{eqnarray}
where the Racah operators $O_{k,q}(x)$ ($2^{k}$-pole operators) satisfy the relations:
\begin{eqnarray}
O_{k,k}(x) &=& \frac{(-1)^{k}}{2^{k}k!}\left[(2k)!\right]^{1/2}\left(S_{+}(x)\right)^{k},\label{ro1}\\
O_{k,q}^{\dagger}(x) &=& (-1)^{q}O_{k,-q}(x),\label{ro2}\\
\left[S_{3}(x),O_{k,q}(y)\right] &=& qO_{k,q}(x)\delta_{xy},\label{ro3}\\
\left[S_{\pm}(x),O_{k,q}(y)\right] &=& \sqrt{k(k+1)-q(q\pm 1)}O_{k,q\pm 1}(x)\delta_{xy},
\nonumber\\\label{ro4}
\end{eqnarray}  
with $S_{\pm}(x)=S_{1}(x)\pm {\rm i}S_{2}(x)$ and $-k\le q\le -k$.\cite{LD} 
$-I_{k}(|x-y|)$ are the $2^{k}$-pole interaction coefficients between sites $x$ and $y$.
Relations between multipole interactions and higher powers of Heisenberg interaction 
are known to be
\begin{eqnarray}
&&\sum_{q=-1}^{1}O_{1,q}(x)O_{1,q}^{\dagger}(y) = \mbox{\boldmath $S$}(x)\cdot\mbox{\boldmath $S$}(y),\\
&&\sum_{q=-2}^{2}O_{2,q}(x)O_{2,q}^{\dagger}(y)\nonumber\\
&&= \frac{3}{2}\left(\mbox{\boldmath $S$}(x)\cdot\mbox{\boldmath $S$}(y)\right)^{2}
+\frac{3}{4}\mbox{\boldmath $S$}(x)\cdot\mbox{\boldmath $S$}(y)-\frac{1}{2}S^{2}(S+1)^{2},
\end{eqnarray}
and for $k\ge 3$ they are given by equations (B.20) and (B.21) in reference \cite{FMH}. 
So $H_{2}$ is written as
\begin{eqnarray}
H_{2}&=&-\sum_{x\ne y\in\Lambda}I_{1}(|x-y|)\sum_{q=-1}^{1}O_{1,q}(x)O_{1,q}^{\dagger}(y)\nonumber\\
&-&\sum_{x\ne y\in\Lambda}I_{2}(|x-y|)\sum_{q=-2}^{2}O_{2,q}(x)O_{2,q}^{\dagger}(y)\label{H2}
\end{eqnarray}
with
\begin{eqnarray}
I_{1}(|x-y|) &=& J_{1}(|x-y|)-\frac{J_{2}(|x-y|)}{2},\\
I_{2}(|x-y|) &=& \frac{2}{3}J_{2}(|x-y|),
\end{eqnarray}
where we have omitted a constant term in the right hand side of equation (\ref{H2}).

Let us also introduce the total spin operator:
\begin{eqnarray}
\mbox{\boldmath $S$}^{\rm tot}
&=& \left(S_{1}^{\rm tot},S_{2}^{\rm tot},S_{3}^{\rm tot}\right)\nonumber\\
&=& \left(\sum_{x\in\Lambda}S_{1}(x),\sum_{x\in\Lambda}S_{2}(x),\sum_{x\in\Lambda}S_{3}(x)\right).
\end{eqnarray}
We easily see continuous symmetry of ${\cal H}_{n}$:
\begin{eqnarray}
\left[S_{1}^{\rm tot},{\cal H}_{n}\right]=\left[S_{2}^{\rm tot},{\cal H}_{n}\right]=\left[S_{3}^{\rm tot},{\cal H}_{n}\right]=0.\label{ri}
\end{eqnarray}
\section{\label{sec3}Models on even numbered rings}
In this section we discuss conditions for establishment of the Lieb-Schupp theorem 
which applies to Hamiltonian (\ref{calH}) on even numbered rings and 
properties of its ground state. 

Before we move forward, let us explain a setup of finite size lattices $\Lambda$ which is needed to establish the Lieb-Schupp theorem. 
Throughout this paper we consider $\Lambda=\Lambda_{\rm L}\cup\Lambda_{\rm R}$ which has an even number of independent sites and $\Lambda$ can be split in two equal parts $\Lambda_{\rm L}$ and $\Lambda_{\rm R}$. 
$\Lambda_{\rm L}$ and $\Lambda_{\rm R}$ are mirror images of one another about a symmetry plane 
without sites which cuts bonds between sites $x\in\Lambda_{\rm L}$ and $y\in\Lambda_{\rm R}$,  
and the collection of these sites is denoted by $\Lambda_{\rm C}$ 
if $\Lambda$ is single even numbered rings. 

In the following we treat the models on just single even numbered rings, i.e., 
closed chains with even number of sites, 
to prepare constructions of the models on two dimensional lattices in section \ref{sec4}.
\subsection{\label{sec3A}Ground state of models on even numbered rings and their reflection symmetry}
Following the above manner let us write Hamiltonian (\ref{calH}) 
on single even numbered rings with $2m$ sites:
\begin{eqnarray}
h_{\rm ring}^{2m}=h_{\rm L}^{m}+h_{\rm R}^{m}+h_{\rm C}^{2m},\label{ring}
\end{eqnarray} 
where
\begin{eqnarray}
h_{\rm L}^{m} &=& -\sum_{x\ne y\in\Lambda_{\rm L}}\sum_{k=1}^{n}\sum_{q=-k}^{k}I_{k}(|x-y|) O_{k,q}(x)O_{k,q}^{\dagger}(y),\nonumber\\ \label{ringL}\\
h_{\rm R}^{m} &=& -\sum_{x\ne y\in\Lambda_{\rm R}}\sum_{k=1}^{n}\sum_{q=-k}^{k}I_{k}(|x-y|) O_{k,q}(x)O_{k,q}^{\dagger}(y),\nonumber\\ \label{ringR}\\
h_{\rm C}^{2m} &=& \sum_{k=1}^{n}(-1)^{k-1}\mbox{\boldmath $O$}_{k}^{\rm L}\cdot{\mbox{\boldmath $O$}_{k}^{\rm R}}^{\dagger},\label{ringC}
\end{eqnarray}
with
\begin{eqnarray}
\mbox{\boldmath $O$}_{k}^{\rm L}
&=& \left(O_{k,k}^{\rm L},O_{k,k-1}^{\rm L},\cdots,O_{k,-k}^{\rm L}\right)\nonumber\\
&=& \sum_{x\in\Lambda_{\rm L}}\alpha_{k}(x)\left(O_{k,k}(x),O_{k,k-1}(x),\cdots,O_{k,-k}(x)\right),\nonumber \\ \label{OL}\\
\mbox{\boldmath $O$}_{k}^{\rm R}
&=& \left(O_{k,k}^{\rm R},O_{k,k-1}^{\rm R},\cdots,O_{k,-k}^{\rm R}\right)\nonumber\\
&=& \sum_{x^{\prime}\in\Lambda_{\rm R}}\alpha_{k}(x^{\prime})\left(O_{k,k}(x^{\prime}),O_{k,k-1}(x^{\prime}),\cdots,O_{k,-k}(x^{\prime})\right).\nonumber \\ \label{OR}
\end{eqnarray}
$h_{\rm L(R)}^{m}$ is a collection of bonds between sites $x,y\in\Lambda_{\rm L(R)}$ 
with $m$ sites.
$h_{\rm C}^{2m}$ contains parallel and crossing bonds between sites $x\in\Lambda_{\rm L}$ and
 $y\in \Lambda_{\rm R}$ which have $2m$ sites. 
Parallel bonds are interactions between $x\in\Lambda_{\rm L}$ 
and $y=x^{\prime}\in\Lambda_{\rm R}$. Here $x^{\prime}$ means a reflection symmetric lattice point of $x$ about the symmetry plane. 
So parallel bonds are perpendicular to the symmetry plane. 
Crossing bonds between $x\in\Lambda_{\rm L}$ and $y\ne x^{\prime}\in\Lambda_{\rm R}$ 
are not perpendicular to it. 
$\mbox{\boldmath $O$}_{k}^{\rm L(R)}$ has $2k+1$ components which are given by, 
for $-k\le q \le k$, $\sum_{x\in\Lambda_{\rm L(R)}}\alpha_{k}(x)O_{k,q}(x)$ 
with real coefficients $\alpha_{k}(x) =\alpha_{k}(x^{\prime})$.
In Hamiltonian (3) interaction coefficients $-I_{k}(|x-y|)$ depend on distance 
between sites $x$ and $y$, but, for the subsequent discussions, 
we consider Hamiltonian (\ref{ring}) containing site-dependent interactions in $h_{\rm C}^{2m}$.

\begin{figure}[t]
\begin{center}
\includegraphics[scale=1]{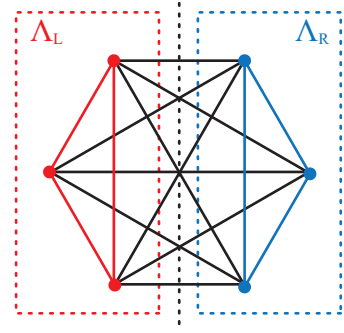}
\caption{Even numbered rings $\Lambda$ split into equal parts $\Lambda_{\rm L}$ and $\Lambda_{\rm R}$ 
which are mirror images of one another with respect to symmetry plane 
denoted by the black dashed line.
As an example of the setup of even numbered rings, $h_{\rm ring}^{6}$ is shown. 
It is decomposed as $h_{\rm L}^{3}$, $h_{\rm R}^{3}$, and $h_{\rm C}^{6}$, 
which limitedly contains red, blue, and black bonds denoted by solid lines, respectively.}
\label{singlehexa}
\end{center}
\end{figure}

To clarify the setup of the even numbered rings, we explain examples of $h_{\rm ring}^{2m}$. In the case of $m=1$ $h_{\rm ring}^{2}$ is just one parallel bond, 
$m=2$ single square with two parallel bonds and two crossing bonds, 
and $m=3$ single hexagon with three parallel bonds and six crossing bonds.  
As one of examples, $h_{\rm ring}^{6}$ is illustrated in FIG. \ref{singlehexa}. 
For $m\ge 4$ they are given by the same manner.

Let us perform unitary transformation on Hamiltonian (\ref{ring}):
\begin{eqnarray}
\widetilde{h}_{\rm ring}^{2m} &=& U^{\dagger}h_{\rm ring}^{2m}U\nonumber\\
&=& h_{\rm L}^{m}+h_{\rm R}^{m}
-\sum_{k=1}^{n}\sum_{q=-k}^{k}O_{k,q}^{\rm L}O_{k,q}^{\rm R},\label{ring1}
\end{eqnarray}
with
\begin{eqnarray}
U = \exp{\left({\rm i}\pi \sum_{x\in\Lambda_{\rm R}}S_{2}(x)\right)},
\end{eqnarray}
where we have used 
\begin{eqnarray}
U^{\dagger}O_{k,q}^{\dagger}(x)U = (-1)^{k}O_{k,q}(x),\label{U}
\end{eqnarray}
for $x \in \Lambda_{\rm R}$. 
The matrix elements of all matrices appearing in $\widetilde{h}_{\rm ring}^{2m}$ are real 
since $O_{k,\pm k}$ are given by the $k$-th power of $S_{\pm}$ with real matrix elements 
and $O_{k,q}$ are generated by the repeated use of commutation relation (\ref{ro4}) 
between $O_{k,\pm k}$ and $S_{\pm}$ which have real matrix elements.

Now we can write a ground state of $\widetilde{h}_{\rm ring}^{2m}$,
\begin{eqnarray}
\widetilde{\psi}=\sum_{\alpha,\beta}c_{\alpha\beta}\psi_{\alpha}^{\rm L}\otimes\psi_{\beta}^{\rm R},
\end{eqnarray}
with real coefficient matrix $c_{\alpha\beta}$,
where $\{\psi_{\alpha}^{\rm L}\}$ form a real orthonormal basis of $S_{3}$ eigenstates 
for the left subsystem and $\{\psi_{\beta}^{\rm R}\}$ are the corresponding states for the right subsystem.
The ground state energy of $\widetilde{h}_{\rm ring}^{2m}$ is given by 
\begin{eqnarray}
\left\langle\widetilde{\psi}\left|\widetilde{h}_{\rm ring}^{2m}\right|\widetilde{\psi}\right\rangle
&=& {\rm Tr}cc^{\dagger}h_{\rm L}^{m}+{\rm Tr}c^{\dagger}ch_{\rm R}^{m}\nonumber\\
&-& \sum_{k=1}^{n}\sum_{q=-k}^{k}{\rm Tr}c^{\dagger}O_{k,q}^{\rm L}cO_{k,q}^{{\rm R}\dagger},\label{e-1}
\end{eqnarray}
with
\begin{eqnarray}
\left(h_{\rm L}^{m}\right)_{\alpha\gamma}
&=& \left\langle\psi_{\alpha}^{\rm L}\left|h_{\rm L}^{m}\right|\psi_{\gamma}^{\rm L}\right\rangle, \\
\left(O_{k,q}^{\rm L}\right)_{\alpha\gamma}
&=& \left\langle\psi_{\alpha}^{\rm L}\left|\sum_{x\in\Lambda_{\rm L}}\alpha_{k}(x)O_{k,q}(x)\right|\psi_{\gamma}^{\rm L}\right\rangle,
\end{eqnarray}
and similarly for $\left(h_{\rm R}^{m}\right)_{\beta\delta}$ and 
$\left(O_{k,q}^{\rm R}\right)_{\beta\delta}$.

Following the arguments in references \cite{lieb-schupp1,lieb-schupp2,schupp,KLS,lieb} 
we set $c\to c^{\dagger}$, then the right hand side of equation (\ref{e-1}) is written as
\begin{eqnarray}
{\rm Tr}c^{\dagger}ch_{\rm L}^{m}+{\rm Tr}cc^{\dagger}h_{\rm R}^{m}
-\sum_{k=1}^{n}\sum_{q=-k}^{k}{\rm Tr}c^{\dagger}O_{k,q}^{\rm R}cO_{k,q}^{{\rm L}\dagger},\label{e-2}
\end{eqnarray}
where we have used equation (\ref{ro2}) and the cyclic property of trace in the third term. 
Here we note that Hamiltonian (\ref{ring1}) is left-right symmetric. 
So we can see that the ground state energy is unchanged and
eigenstates with coefficient matrices $c^{\dagger}$ and $c+c^{\dagger}$ are also ground states. 
There exists at least one ground state with Hermite coefficient matrix. 
Hermite coefficient matrix can be diagonalized and then the third term in the right-hand side of equation (\ref{e-1}) are written as 
\begin{eqnarray}
-\sum_{k=1}^{n}\sum_{q=-k}^{k}\sum_{l,m}c_{ll}c_{mm} \left|\left(O_{k,q}\right)_{lm}\right|^{2}, \label{cross1}
\end{eqnarray}
in the diagonal basis of $c$. This expression is bounded below by
\begin{eqnarray}
-\sum_{k=1}^{n}\sum_{q=-k}^{k}\sum_{l,m}\left|c_{ll}\right|\left|c_{mm}\right| \left|\left(O_{k,q}\right)_{lm}\right|^{2}. \label{cross2}
\end{eqnarray} 
So we can confirm that an eigenstate with positive semidefinite coefficient matrix $|c|=\sqrt{c^{2}}$ is a ground state of $\widetilde{h}_{\rm ring}^{2m}$.
\subsection{\label{sec3B}Singlet Ground States}
In this subsection, at first, we confirm that a ground state of $h_{\rm ring}^{2m}$, 
i.e., $\psi=U\widetilde{\psi}$ with positive semidefinite coefficient matrix $|c|$ has $S^{\rm tot}=0$. 

Let us introduce a tensor product of spin singlet state:
\begin{eqnarray}
\psi_{0}=\bigotimes_{x\in\Lambda_{\rm L}}\sum_{M=-S}^{S}(-1)^{S-M}\left|S,M\right\rangle_{x}\otimes\left|S,-M\right\rangle_{x^{\prime}},
\end{eqnarray}
where $M$ is eigenvalues of $S_{3}(x)$ and $x\in\Lambda_{\rm L}$ 
counts every pair $x$ and its reflection symmetric point $x^{\prime}$. 
Ground state $\widetilde{\psi}$ with $|c|$ is written as
\begin{eqnarray}
\widetilde{\psi}=\sum_{\alpha}\left|c\right|_{\alpha\alpha}\psi_{\alpha}^{\rm L}\otimes\psi_{\alpha}^{\rm R}.
\end{eqnarray}
We can easily see
\begin{eqnarray}
\left\langle\psi_{0}|\psi\right\rangle=\left\langle\psi_{0}\left| U\right|\widetilde{\psi}\right\rangle={\rm Tr}|c|>0.
\end{eqnarray}
Since $\psi_{0}$ has $S^{\rm tot}=0$ and ${S}^{\rm tot}$ is a good quantum number, 
ground state $\psi$ must take $S^{\rm tot}=0$. 
Thus we can find that there exists at least one ground state with $S^{\rm tot}=0$. 
This result makes strong in the following argument.

Next, we show that all ground states of $h_{\rm ring}^{2m}$ have $S^{\rm tot}=0$ 
even if the ground state is degenerate. 
Let $b(x)$ be a real valued function of site $x$. 
Now we consider the unitarily transformed Hamiltonian under site-dependent field given by
\begin{eqnarray}
\widetilde{h}_{\rm ring}^{2m}(b) &=& h_{\rm L}^{m}+h_{\rm R}^{m}\nonumber\\
&-& \left[\sum_{k\ne l}^{n}\sum_{q=-k}^{k}O_{k,q}^{\rm L}O_{k,q}^{\rm R}
+\sum_{q\ne0}O_{l,q}^{\rm L}O_{l,q}^{\rm R}\right.\nonumber\\
&+&\left(O_{l,0}^{\rm L}-\sum_{x\in\Lambda_{\rm L}}b(x)\right)
\left(O_{l,0}^{\rm R}-\sum_{x^{\prime}\in\Lambda_{\rm R}}b(x^{\prime})\right)\nonumber\\
&-&\frac{1}{2}\left(O_{l,0}^{\rm L}-\sum_{x\in\Lambda_{\rm L}}b(x)\right)^{2}
+\frac{1}{2}\left(O_{l,0}^{\rm L}\right)^{2}\nonumber\\
&-&\left.\frac{1}{2}\left(O_{l,0}^{\rm R}-\sum_{x^{\prime}\in\Lambda_{\rm R}}b(x^{\prime})\right)^{2}
+\frac{1}{2}\left(O_{l,0}^{\rm R}\right)^{2}\right]
\label{hb}
\end{eqnarray}
Here we note that $\widetilde{h}_{\rm ring}^{2m}(0)=U^{\dagger}h_{\rm ring}^{2m}U$. 
Following the argument of Kennedy-Lieb-Shastry with a trace inequality\cite{KLS,TTI}, we get
\begin{eqnarray}
e_{\rm ring}^{2m}(b)\ge e_{\rm ring}^{2m}(0),\label{energy}
\end{eqnarray}
concerning for the ground state energy of $\widetilde{h}_{\rm ring}^{2m}(b)$. 
It is required for establishment of this inequality to satisfy the conditions:
the matrix elements of the matrix representations of $O_{k,q}^{\rm L}$ and $O_{k,q}^{\rm R}$ 
are real and the coefficients of all interaction terms $O_{k,q}^{\rm L}O_{k,q}^{\rm R}$ are negative. 

When we choose
\begin{eqnarray}
b(x)=
\left\{ \begin{array}{ll}
    b &x \in \Lambda_{\rm L},\\
    -b& x\in\Lambda_{\rm R}
  \end{array} \right.
\end{eqnarray}
in equation (\ref{hb}), it becomes
\begin{eqnarray}
\widetilde{h}_{\rm ring}^{2m}(b)=\widetilde{h}_{\rm ring}^{2m}
-2mb\left(O_{l,0}^{\rm L}-O_{l,0}^{\rm R}\right)+2m^{2}b^{2}.
\end{eqnarray}
Let $\widetilde{\psi}(b)$ be a ground state of $\widetilde{h}_{\rm ring}^{2m}(b)$. By the variational principle and inequality (\ref{energy}), we have
\begin{eqnarray}
\left\langle \widetilde{\psi}(0)\left|\widetilde{h}_{\rm ring}^{2m}(b)\right|\widetilde{\psi}(0)\right\rangle
&\ge&\left\langle \widetilde{\psi}(b)\left|\widetilde{h}_{\rm ring}^{2m}(b)\right|\widetilde{\psi}(b)\right\rangle\nonumber\\
&=&e_{\rm ring}^{2m}(b)\ge e_{\rm ring}^{2m}(0),
\end{eqnarray}
which leads to
\begin{eqnarray}
-mb\left\langle \widetilde{\psi}(0)\left|O_{l,0}^{\rm L}-O_{l,0}^{\rm R}\right|\widetilde{\psi}(0)\right\rangle+m^{2}b^{2}\ge 0.
\end{eqnarray}
This result is independent of value of $b$.
In order to establish this inequality for arbitrary value of $b$ it must be 
\begin{eqnarray}
\left\langle \widetilde{\psi}(0)\left|O_{l,0}^{\rm L}-O_{l,0}^{\rm R}\right|\widetilde{\psi}(0)\right\rangle=0.
\end{eqnarray}
Noting that $U\widetilde{\psi}(0)=\psi(0)$ is $\psi$, then we get
\begin{eqnarray}
&&\left\langle \psi\left|\sum_{x\in\Lambda_{\rm L}}\alpha_{l}(x)O_{l,0}(x)\right.\right.\nonumber\\
&&\left.\left.-(-1)^{l}\sum_{x^{\prime}\in\Lambda_{\rm R}}\alpha_{l}(x^{\prime})O_{l,0}(x^{\prime})\right|\psi\right\rangle
=0.\label{icerule}
\end{eqnarray}
Setting $l=1$, we can see $\left\langle \psi\left|S_{3}^{\rm tot}\right|\psi\right\rangle=0$ 
since the above equation must establish arbitrary values of $\alpha_{1}(x)$ for all $x$. 
By the rotational invariance of $h_{\rm ring}^{2m}$ as is shown in equation (\ref{ri}),
this result also holds for $S_{1}^{\rm tot}$ and $S_{2}^{\rm tot}$. So it concludes that the all ground states have $S^{\rm tot}=0$.
\section{\label{sec4}Constructions of models on lattices 
with local Hamiltonian on even numbered rings}
In the previous section 
we have showed that all ground states of $h^{2m}_{\rm ring}$ 
on even numbered rings have $S^{\rm tot}=0$.
In this section we consider constructions of global Hamiltonian on two dimensional lattices 
with local Hamiltonian $h^{2m}_{\rm ring}$. 
Here we suppose that whole lattices are constructed with even numbered rings, 
such as square lattice, honeycomb lattice, 
1/5-depleted lattice (CaV$_{4}$O$_{9}$).\cite{depleted} 
1/5-depleted lattice consists of squares and octagons. 

In subsection \ref{sec4A}, we show that all ground states of global Hamiltonian 
with site-dependent interactions possesses $S^{\rm tot}=0$. 
In subsection \ref{sec4B}, 
we consider models on lattices without crossing bonds 
as a special case of global Hamiltonian in subsection \ref{sec4A} and 
determine conditions realizing spatially isotropic interactions. 
In subsection \ref{sec4C} and \ref{sec4D}, 
as examples of lattices with crossing bonds in this framework, 
we perform the same procedure in subsection \ref{sec4B} 
in the case of square and honeycomb lattices with crossing bonds. 
\subsection{\label{sec4A}Generalized lattices} 
In section \ref{sec3} we have treated $h_{\rm ring}^{2m}$ on even numbered rings  
and have showed that their ground states possess $S^{\rm tot}=0$. 
It is straightforward to prove that ground states of global Hamiltonian on generalized lattices 
have the same result. 
In this subsection we shortly explain it as follows.

Let us write global Hamiltonian:
\begin{eqnarray}
{\cal H}={\cal H}_{\rm L}+{\cal H}_{\rm R}+{\cal H}_{\rm C}.\label{wh}
\end{eqnarray}
${\cal H}_{\rm C}$ is constructed with translated copies of local Hamiltonian 
$\sum_{j}^{m}h_{\rm C}^{2j}$ on an even numbered ring with $2m$ sites 
in the direction parallel to the symmetry plane,\cite{bc} i.e., 
\begin{eqnarray}
{\cal H}_{\rm C}=\sum_{\Lambda_{\rm C}}\sum_{j}^{m}h_{\rm C}^{2j},\label{whc}
\end{eqnarray}
where the summation for $j$ is taken if it is necessary and the same applies hereinafter.
The sites belongings to $\Lambda_{\rm C}$ are denoted by the collection of sites 
which are translated copies of sites belonging to an even numbered ring 
such as single square and single hexagon in the direction parallel to the symmetry plane(see FIG.2).

Let $h_{\rm ring}^{2m}=h_{\rm L}^{m}+h_{\rm R}^{m}+\sum_{j}^{m}h_{\rm C}^{2j}$. 
${\cal H}_{\rm L(R)}$ consists of the collection of translated copies of $h_{\rm ring}^{2m}$ 
on $\Lambda_{\rm L(R)}$ and 
the collection of bonds in $h_{\rm L(R)}^{m}$ on ${\partial}\Lambda_{\rm L(R)}$. 
The sites belonging to $\partial{\Lambda}_{\rm L(R)}$ 
is denoted by $\Lambda_{\rm L(R)}\cap\Lambda_{\rm C}$
(the sites belonging to $\Lambda_{\rm C}={\partial}\Lambda_{\rm L}\cup{\partial}\Lambda_{\rm R}$).  
Then, global Hamiltonian on $\Lambda_{\rm L(R)}$ is written as 
\begin{eqnarray}
{\cal H}_{\rm L(R)}=\sum_{\Lambda_{\rm L(R)}}h_{\rm ring}^{2m}+\sum_{\partial\Lambda_{\rm L(R)}}h_{\rm L(R)}^{m}.\label{ghlr}
\end{eqnarray}
The second term in this Hamiltonian is omitted if global Hamiltonian is constructed with bond sharing even numbered rings.  
These operations should be realized to construct the global Hamiltonian 
on the two dimensional lattices $\Lambda=\Lambda_{\rm L}\cup\Lambda_{\rm R}$.
We recall that $\Lambda_{\rm L}$ and $\Lambda_{\rm R}$ are equal parts and 
$\Lambda$ has reflection symmetry about the symmetry plane. 
Here we use `generalized' lattices in the sense that global Hamiltonian (\ref{wh}) 
is constructed with $h_{\rm C}^{2j}$ containing site-dependent interactions 
in spite that ${\cal H}_{n}$ has spatially isotropic interactions.  

We can easily see that global Hamiltonian (\ref{wh}) also holds the same results in section \ref{sec3}. 
Roughly speaking,  
main differences are that, in equations (\ref{hb}) and (\ref{energy}),  
$h_{\rm L(R)}^{m}$ and local ground state $\psi$ 
are replaced by ${\cal H}_{\rm L(R)}$ and global ground state 
$\Psi=\sum_{\alpha,\beta}C_{\alpha\beta}\Psi_{\alpha}^{\rm L}\otimes\Psi_{\beta}^{\rm R}$ 
and $\sum_{\Lambda_{\rm c}}\sum_{j}^{m}$ appears 
in the terms of parallel and crossing bonds on $\Lambda_{\rm C}$. 
Through the same procedure in subsection \ref{sec3B} 
we can see similar equation 
with respect to $\Lambda_{\rm C}(=\partial \Lambda_{\rm L}\cup\partial \Lambda_{\rm R})$ 
to equation (\ref{icerule}) as follows.\cite{note1}
\begin{eqnarray}
&&\left\langle\Psi\left|\sum_{x\in\partial\Lambda_{\rm L}}\alpha(x)O_{l,0}(x)\right.\right.\nonumber\\
&&\left.\left.-(-1)^{l}\sum_{x^{\prime}\in\partial\Lambda_{\rm R}}\alpha(x^{\prime})O_{l,0}(x^{\prime})\right|\Psi\right\rangle=0\label{ghb1}
\end{eqnarray}
and
\begin{eqnarray}
\left\langle\Psi\left|\sum_{x\in\partial\Lambda_{\rm L}}O_{l,0}(x)
-(-1)^{l}\sum_{x^{\prime}\in\partial\Lambda_{\rm R}}O_{l,0}(x^{\prime})\right|\Psi\right\rangle=0.\label{ghb2}\nonumber\\
\end{eqnarray}
Let us set $l=1$ and impose a periodic boundary condition in the direction 
perpendicular to the symmetry plane, 
we conclude that all ground states of global Hamiltonian have $S^{\rm tot}=0$ 
if whole lattices can be constructed with translated copies of $\Lambda_{\rm C}$.
Otherwise, we need to set other symmetry planes and 
impose periodic boundary conditions in the directions perpendicular to those symmetry planes.
\subsection{\label{sec4B}Lattices without crossing bonds}
In the case of bipartite lattices, 
whole lattices are constructed with even numbered rings without crossing bonds. 
so we can construct global Hamiltonian with $h^{2}_{\rm ring} (m=1)$, 
i.e., nearest neighbor pairs $\langle x,y \rangle$ only.
This simplest model of Hamiltonian (\ref{wh}) is written as 
\begin{eqnarray}
&&\sum_{\langle x,y \rangle\in\Lambda}h^{2}_{\rm ring}\nonumber\\
&&=\sum_{\langle x,y \rangle\in\Lambda}\sum_{k=1}^{n}(-1)^{k-1}\alpha_{k}(x)^{2}
\sum_{q=-k}^{k}O_{k,q}(x)O_{k,q}^{\dagger}(y).\label{gh2}
\end{eqnarray}
Interaction coefficients in the right hand side of this equation 
are correspond to $-I_{k}(1)$ and we see $(-1)^{k}I_{k}(1)\ge 0$ for each $k$.
\subsection{\label{sec4C}Square lattices with crossing bonds}
\begin{figure}[t]
\begin{center}
\includegraphics[scale=1]{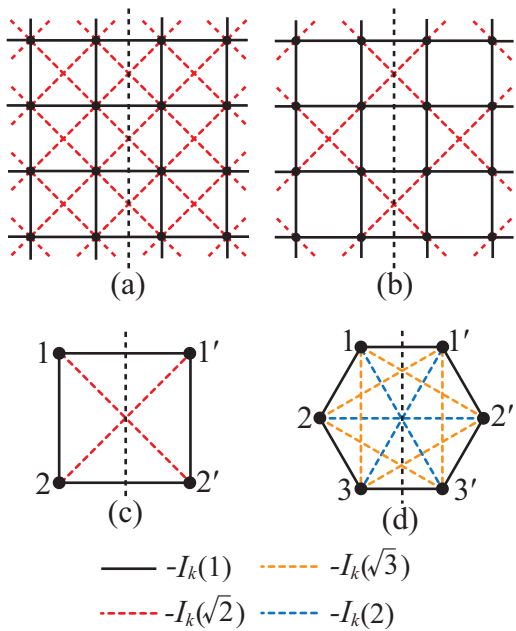}
\caption{(a) Square lattice with crossing bonds. (b) Checkerboard type. 
(c) Local Hamiltonian (\ref{h4}) on single square with crossing bonds. 
Black solid and red dashed lines represent nearest neighbor interaction $-I_{k}(1)$ 
and second neighbor interaction $-I_{k}(\sqrt{2})$, respectively. 
(d) Local Hamiltonian (\ref{h6}) on single hexagon with crossing bonds. 
Black solid, orange dashed, 
and blue dashed lines represent nearest neighbor interaction $-I_{k}(1)$, 
second neighbor interaction $-I_{k}(\sqrt{3})$, and third neighbor interaction $-I_{k}(2)$, respectively. 
Black dashed lines in FIGs (a)-(d) represent reflection symmetric plane. 
}
\end{center}
\end{figure}
In references \cite{lieb-schupp1,lieb-schupp2,KK,woj} 
the antiferromagnetic Heisenberg models 
with next nearest neighbor interactions on the square lattice and 
its checkerboard type were discussed. 
In this subsection, following these previous studies 
we treat the case of Hamiltonian (\ref{calH}).  

Now let us consider constructions of global Hamiltonian 
on the square lattice with $h_{\rm ring}^{4} (m=2)$. 
$h_{\rm ring}^{4}$ is local Hamiltonian on single square with crossing bonds (see FIG.3-(c)).
To determine the condition realizing spatially isotropic interactions of the models we write 
\begin{eqnarray}
&&h_{\rm ring}^{4} 
- \sum_{k=1}^{n}\sum_{q=-k}^{k}(-1)^{k-1}
\sum_{x=1}^{2}\alpha_{k}(x)^{2}O_{k,q}(x)O_{k,q}^{\dagger}(x^{\prime})\nonumber\\
&&-\sum_{k=1}^{n}\sum_{q=-k}^{k}I_{k}(1)\sum_{x=1}^{2}
O_{k,q}(x)O_{k,q}^{\dagger}(x^{\prime})\nonumber\\
&&=h_{\rm ring}^{4}+\sum_{k=1}^{n}\sum_{q=-k}^{k}\sum_{x=1}^{2}\left[-I_{k}(1)\right.\nonumber\\
&&\left.-(-1)^{k-1}\alpha_{k}(x)^{2}\right]O_{k,q}(x)O_{k,q}^{\dagger}(x^{\prime}),
\label{h4}
\end{eqnarray}
where $-I_{k}(1)$ are denoted by coefficients of 
the first neighbor $2^{k}$-pole interactions. 
To explain the lattice structures and their spatial isotropy, they are illustrated in FIGs 2-(a),(b),(c). 
Noting that the factor $(-1)^{k}$ appears from each term in the right hand side of this equation 
through the unitary transformation, 
then it can be seen as a combination of $h_{\rm ring}^{4}$ and $h_{\rm ring}^{2}$ if interaction coefficients of these local Hamiltonian satisfy
\begin{eqnarray}
-(-1)^{k}I_{k}(1) + \alpha_{k}(1)^{2}\le 0,\\
-(-1)^{k}I_{k}(1) + \alpha_{k}(2)^{2}\le 0,
\end{eqnarray}
for each $k$ and we get
\begin{eqnarray}
&&(-1)^{k}I_{k}(1) \ge 0,\label{I1} \\
&&-I_{k}(1) \le -I_{k}(\sqrt{2})\le I_{k}(1), \label{I2}
\end{eqnarray}
with second neighbor interaction coefficients $-I_{k}(\sqrt{2})=(-1)^{k-1}\alpha_{k}(1)\alpha_{k}(2)$.

When global Hamiltonian is constructed with local Hamiltonian (\ref{h4}) 
on bond sharing squares as in FIG 2-(a), 
the results in subsection \ref{sec4A} hold if $-I_{k}(1)/2\le I_{k}(\sqrt{2})\le I_{k}(1)/2$ 
and $(-1)^{k}I_{k}(1)\ge 0$. 
But its checkerboard type is site sharing as in FIG 2-(b).
So condition $-I_{k}(1)\le I_{k}(\sqrt{2})/2\le I_{k}(1)$ is replaced by $-I_{k}(1)\le I_{k}(\sqrt{2})\le I_{k}(1)$. 
\subsection{\label{sec4D}Honeycomb lattice with crossing bonds}
In this subsection 
we treat models on the honeycomb lattice with crossing bonds which is constructed with 
translated copies of local Hamiltonian on single hexagon (FIG 2-(d)). 
Let us consider global Hamiltonian with ${\cal H}_{\rm C}$ given by
\begin{eqnarray}
\sum_{\Lambda_{\rm C}}\sum_{j=2}^{3}h_{\rm C}^{2j}
&=& \sum_{\Lambda_{\rm C}}\left(h_{\rm C}^{6}+h_{\rm C}^{4}\right). 
\end{eqnarray}
In FIG. 2-(d) $h_{\rm C}^{4}$ is local Hamiltonian on a rectangular formed by 
four sites $1,1^{\prime},3,3^{\prime}$ except dashed orange bonds.
So inner product of $\mbox{\boldmath $O$}_{k}^{\rm L}$ 
and ${\mbox{\boldmath $O$}_{k}^{\rm R}}^{\dagger}$ in $h_{\rm C}^{4}$ is defined by  
\begin{eqnarray}
\mbox{\boldmath $O$}_{k}^{\rm L}
&=& \sum_{x=1,3}\beta_{k}(x)\left(O_{k,k}(x),O_{k,k-1}(x),\cdots,O_{k,-k}(x)\right)\nonumber\\
\end{eqnarray}
and ${\mbox{\boldmath $O$}_{k}^{\rm R}}$ with $\beta_{k}(x^{\prime})$ 
on sites $1^{\prime},3^{\prime}$.

Following the previous subsection let us write the local Hamiltonian:
\begin{eqnarray}
&&h_{\rm ring}^{6}-\sum_{k=1}^{n}\sum_{q=-k}^{k}
(-1)^{k-1}\sum_{x=1}^{3}\alpha_{k}(x)^{2}O_{k,q}(x)O_{k,q}^{\dagger}(x^{\prime})\nonumber\\
&&-\sum_{k=1}^{n}\sum_{q=-k}^{k}\sum_{x=1}^{3}I_{k}(|x-x^{\prime}|)O_{k,q}(x)O_{k,q}^{\dagger}(x^{\prime})\nonumber\\
&&+h_{\rm C}^{4}-\sum_{k=1}^{n}\sum_{q=-k}^{k}(-1)^{k-1}\sum_{x=1,3}\beta_{k}(x)^{2}O_{k,q}(x)O_{k,q}^{\dagger}(x^{\prime})\nonumber\\
&&=h_{\rm ring}^{6}+h_{\rm C}^{4}\nonumber\\
&&+\sum_{k=1}^{n}\sum_{q=-k}^{k}\left\{\left[-I_{k}(2)-(-1)^{k-1}\alpha_{k}(2)^{2}\right]O_{k,q}(2)O_{k,q}^{\dagger}(2^{\prime})\right.\nonumber\\
&&+\sum_{x=1,3}\left[-I_{k}(1)-(-1)^{k-1}\alpha_{k}(x)^{2}\right.\nonumber\\
&&\left.\left.-(-1)^{k-1}\beta_{k}(x)^{2}\right]O_{k,q}(x)O_{k,q}^{\dagger}(x^{\prime})\right\}
,\label{h6}
\end{eqnarray}
where $-I_{k}(1)$, $-I_{k}(\sqrt{3})$, and $-I_{k}(2)$ are denoted by first, second, and third neighbor interaction coefficients of $2^{k}$-pole interactions, respectively, as illustrated in FIG. 2-(d).
Similar to the previous subsection, 
right hand side of this equation can be seen as a combination of $h_{\rm ring}^{6}$ 
and $h_{\rm C}^{4}$ along with $h_{\rm ring}^{2}$ and its analogue if
\begin{eqnarray}
&&-(-1)^{k}I_{k}(2) +\alpha_{k}(2)^{2} \le 0,\\
&&-(-1)^{k}I_{k}(1)+\alpha_{k}(1)^{2}+\beta_{k}(1)^{2} \le 0,\\
&&-(-1)^{k}I_{k}(1)+\alpha_{k}(3)^{2}+\beta_{k}(3)^{2} \le 0,
\end{eqnarray}
and conditions on interaction parameters satisfying spatial isotropy in $\Lambda$ are given by
\begin{eqnarray}
&&-I_{k}(\sqrt{3}) = (-1)^{k-1}\alpha_{k}(1)\alpha_{k}(2)= (-1)^{k-1}\alpha_{k}(2)\alpha_{k}(3),\nonumber\\\\
&&-I_{k}(2)=(-1)^{k-1}\left[\alpha_{k}(1)\alpha_{k}(3)+\beta_{k}(1)\beta_{k}(3)\right],
\end{eqnarray} 
$\alpha_{k}(1)=\alpha_{k}(3)$, and $\beta_{k}(1)=\beta_{k}(3)$ for each $k$. From these equations and inequalities we get
\begin{eqnarray}
\frac{1}{2}\ge \frac{I_{k}(2)}{I_{k}(1)}\ge 
-\frac{1}{4}+\sqrt{\frac{1}{16}+2\left(\frac{I_{k}(\sqrt{3})}{I_{k}(1)}\right)^{2}},\label{hineq1}\\
(-1)^{k}\frac{I_{k}(1)}{2}\ge (-1)^{k}I_{k}(2)\ge 0,\label{hineq2}
\end{eqnarray}
where we have replaced $I_{k}(1)$ by $I_{k}(1)/2$ since global Hamiltonian is constructed with local Hamiltonian (\ref{h6}) on bond sharing hexagons.

\section{\label{sec5}Summary and discussions}
We have discussed the Heisenberg models 
with higher order interactions or multipole interactions on finite lattices with reflection symmetry
written as in the form of Hamiltonian (\ref{ring}) or (\ref{wh})
and have found that there exists at least one ground state with $S^{\rm tot}=0$. 
Moreover imposing a periodic boundary condition 
in the direction perpendicular to the symmetry plane, 
we have confirmed that the all ground states possess $S^{\rm tot}=0$ 
even if the ground state is degenerate. 
These results are a straightforward extension of the Lieb-Schupp theorem to these models. 

For establishment of the results in subsections \ref{sec3B} and \ref{sec4A} 
we did not put any restrictions on signs or values of interaction coefficients $-I_{k}(|x-y|)$ 
of $h_{\rm L}^{m}$ and $h_{\rm R}^{m}$ 
except their reflection symmetry (see Hamiltonian (\ref{ring})-(\ref{ringC}) and (\ref{wh})-(\ref{ghlr})). 
These coefficients are not essential to our results. 
In subsections \ref{sec4B}, \ref{sec4C}, and \ref{sec4D} they were determined 
by the conditions in order that ${\cal H}_{n}$ possesses spatially isotropic interactions as in FIG. 2. 
On the other hand restrictions on interaction coefficients of parallel bonds and crossing bonds 
in $h_{\rm C}^{2m}$ come from 
the establishment of inequality (\ref{energy}) and similar inequalities for the models in this paper.\cite{note1} 
Therefore we have no idea for improvement of these restrictions. 

In this section, we summarize and discuss the results in section \ref{sec3} and \ref{sec4} 
which are divided into the models on lattices with and without crossing bonds. 
\subsection{Lattices without crossing bonds}
In this case, 
models are constructed with local Hamiltonian $h_{\rm ring}^{2}$. 
Typical examples of the whole lattice are bipartite lattices such as hypercubic lattice, honeycomb lattice, and 1/5-depleted lattice. 
In the case of the 1/5-depleted lattice we should set the symmetry plane 
which intersects octagons and squares.  
The results hold if $(-1)^{k}I_{k}(1)\ge 0$ for each $k$. 

In the following we shall explain comparisons with the results of the Marshall-Lieb-Mattis type argument. 
$H_{1}={\cal H}_{1}$ with $-J_{1}(1)=-I_{1}(1)>0$ is the spin-$S$ antiferromagnetic Heisenberg model. 
The Marshall-Lieb-Mattis theorem assures that its ground state is unique and has $S^{\rm tot}=0$. 
So the result given by the Lieb-Schupp theorem is completely covered by 
the Marshall-Lieb-Mattis theorem with uniqueness of the ground state.

${\cal H}_{2}$ is equivalent to bilinear $-J_{1}(1) (=-I_{1}(1)-3I_{2}(1)/4)$ 
and biquadratic $-J_{2}(1)(=-3I_{2}(1)/2)$ exchange interaction model $H_{2}$.
From the previous studies of the Marshall-Lieb-Mattis type argument it was known that 
the same results hold for $H_{2}$ with 
$S=1$ in the region  $J_{2}(1)>J_{1}(1),J_{2}(1)\ge 0(J_{1}(1)\ne 0)$ 
and with $S>1$ in the region $0\le J_{2}(1)\le -J_{1}(1)/2S(S-1)$.\cite{munro,parkinson,tanaka}
On the other hand, our results based on the Lieb-Schupp theorem show that
all ground states have $S^{\rm tot}=0$ 
in the region $J_{1}(1)\ge 2J_{2}(1), J_{2}(1)\ge 0$ for any $S$. 
For $S=1$ purely biquadratic interaction model ($J_{1}(1)=0,J_{2}(1)>0$) 
satisfies SU(3) symmetry and its ground state is degenerate. 
Our results can conclude that all degenerate ground states possess $S^{\rm tot}=0$. 
In the case of $S>1$ our results extend the region 
which one can conclude ground states with $S^{\rm tot}=0$. 
These results are summarized in FIG. 3.

For ${\cal H}_{n}$ with $n>2$, our results also hold if $(-1)^{k}I_{k}(1)\ge 0$ for each $k$. 
As far as we know, the results of the Marshall-Lieb-Mattis type argument does not exist.

Our study is concerned with models on finite lattices with reflection symmetry 
and their ground states possess $S^{\rm tot}=0$, 
but in infinite volume limit continuous symmetry breaking may occur. 
The antiferromagnetic Heisenberg model on bipartite lattices in two or more dimensions 
is known to exhibit N\'{e}el long range order in its ground state. 
For the $d$-dimensional hypercubic lattice, 
it was rigorously proved in $d\ge 3$ for any $S$ and in $d=2$ for $S\ge 1$\cite{KLS,DLS,NF}, 
and for the honeycomb lattice for $S\ge 3/2$\cite{aklt}. 
Ground state phase diagram of $S=1$ bilinear-biquadratic model ($H_{2}$) with $J_{2}(1)>0$ 
on the square or the simple cubic lattice is considered as follows.\cite{QS1,QS2,QS3,QS4} 
The region $J_{2}(1)>J_{1}(1)>0$ is the ferroquadrupole (spin nematic) phase, 
$J_{1}(1)<0$ the N\'{e}el ordered phase, 
and $0<J_{2}(1)<J_{2}(1)$ the ferromagnetic phase, 
which are illustrated in FIG. 3-(a). 
In our parameter region there exist N\'{e}el long range order and ferroquadrupole long ranger order 
in infinite volume ground state,  
which were rigorously proved in parts of these parameter region.\cite{TTI,U,Lee} 
These proofs were given by the method of infrared bounds 
whose key inequality is upper bound on the Fourier transformed correlation function 
in whole momentum space which is derived from inequality (\ref{energy}) and similar ones 
for the ground state energy or analogous inequality for the partition function.
\begin{figure}[t]
\begin{center}
\includegraphics[scale=0.8]{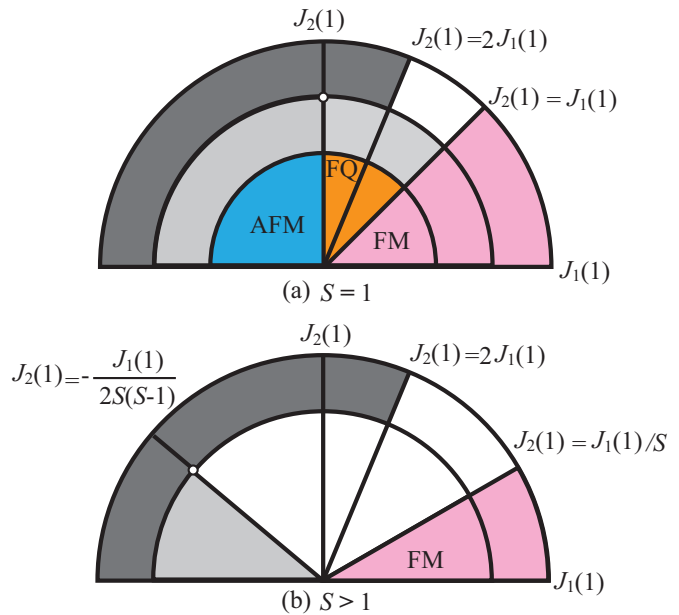}
\caption{(a) $S=1$ and (b) $S>1$ ground state phase diagram of 
the bilinear-biquadratic exchange Hamiltonian on finite lattices without crossing bonds.
Our results based on the Lieb-Schupp theorem mean that all ground states have $S^{\rm tot}=0$ in the dark gray region. 
The results of the Marshall-Lieb-Mattis type arguments mean that 
unique ground state has $S^{\rm tot}=0$ in the light gray region. 
Three colored regions in inner semicircle in FIG. (a) are expected phase diagram of $d\ge 2$ dimensional hyper cubic lattice in infinite volume limit. Pink, orange, and blue region is ferromagnetic (FM), ferroquadrupole (FQ), and N\'eel ordered (AFM) phase, respectively. }
\end{center}
\end{figure}

\subsection{Lattices with crossing bonds}
In this case, models are constructed with translated copies of $h_{\rm ring}^{2m}$ with $m>1$ 
on rings. These rings have crossing bonds. 
In subsection \ref{sec4A} we have proved that 
global Hamiltonian (\ref{wh}) holds the same results in subsection \ref{sec3B}.
These models have site-dependent interactions. 
So in subsections \ref{sec4C} and \ref{sec4D} we considered conditions 
which Hamiltonian (\ref{wh}) possesses spatially isotropic interactions 
on the square and the honeycomb lattices with crossing bonds as in FIG. 2. 
For the non-checkerboard square lattice it is realized in the region
$-I_{k}(1)/2 \le I_{k}(\sqrt{2}) \le I_{k}(1)/2$ 
and checkerboard type $-I_{k}(1) \le I_{k}(\sqrt{2}) \le I_{k}(1)$ for each $k$. 
The result for the honeycomb lattice is given by inequalities (\ref{hineq1}) (\ref{hineq2}), 
and setting $|I_{k}(1)|=1$ it is illustrated in FIG. 4. 

Only about the frustrated antiferromagnetic Heisenberg model 
on the square and honeycomb lattice with crossing bonds, 
we explain relation between our results on finite lattices and 
the results of various theoretical studies on infinite lattices.
There exist detailed reviews of these models on square lattices in the paper \cite{J1-J2} (see also references therein). 
In infinite volume limit, within our parameter region, 
the ground state phase diagram expected to be valid is summarized as follows.
For the non-checkerboard (checkerboard) square lattice, 
in the case of $S=1/2$, 
the region $0 (0)\le J_{1}(\sqrt{2})/J_{1}(1) \lesssim 0.4(0.8)$  
is the N\'{e}el ordered phase and 
$0.4(0.8) \lesssim J_{1}(\sqrt{2})/J_{1}(1)\le 0.5(1)$ the quantum paramagnetic phase 
without magnetic long range order, 
and in $S=1$, $0(0)\le J_{1}(\sqrt{2})/J_{1}(1) \le 0.5(1)$ the N\'{e}el ordered phase. 
In the case of $S=1$ and non-checkerboard type, by using the method of infrared bounds, 
the existence of N\'{e}el long range order 
was rigorously proved in small $J_{1}(\sqrt{2})/J_{1}(1)$.\cite{KK}   
Ground state phase diagram of $S=1/2$ on the honeycomb lattice (non-checkerboard type) 
was obtained in reference \cite{honeycomb1,honeycomb2}. 
Our parameter region also seems to be contained in the N\'{e}el ordered phase and the quantum paramagnetic phase.
\begin{figure}[t]
\begin{center}
\includegraphics[scale=0.8]{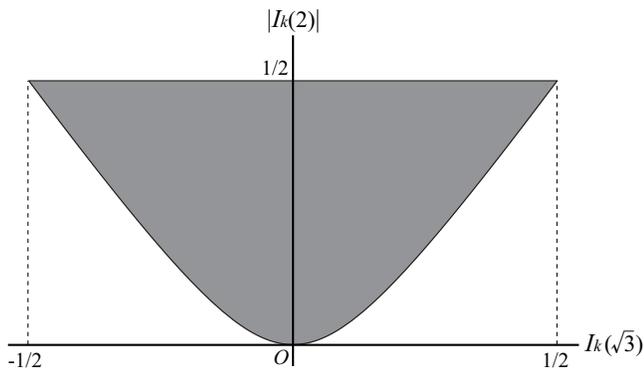}
\caption{Ground states of ${\cal H}_{n}$ on honeycomb lattice with crossing bonds 
have $S^{\rm tot}=0$ in the gray parameter region for each $k$ 
which is determined by inequalities (\ref{hineq1}) and (\ref{hineq2}). 
$I_{k}(1)$ is scaled as $-1$ if $k$ is odd, otherwise $I_{k}(1)=1$, 
and $-1/2\le I_{k}(2)\le 0$ if $k$ is odd, otherwise $0\le I_{k}(2)\le 1/2$.}
\end{center}
\end{figure}

In this paper we focus on one dimensional rings and two dimensional lattices. 
Application of these results to three dimensional lattices can be easily extended.
In that case, as a simplest example, 
we can consider local Hamiltonian on single cubes with crossing bonds 
and it should be written as $h_{\rm box}^{8}$. 

Lieb-Schupp called FIG. 2-(b) pyrochlore checkerboard 
since it is a two dimensional projection of a three dimensional pyrochlore lattice.
But this framework is not applicable to the pyrochlore lattice unfortunately, 
because it lacks reflection symmetry.
They also called equation (\ref{icerule}) with $l=1$ quantum analogue of ice rule 
in the context of the correspondence between Ising like ferromagnet 
with crystal field anisotropy on the pyrochlore lattice and 
configuration of four hydrogen atoms around an oxygen atom in ice.\cite{lieb-schupp1,lieb-schupp2} 
In that point of view equations (\ref{icerule}) and (\ref{ghb2}) are generalization of ice rule to 
any even numbered frustrated units and $2^{2l-1}$ pole moment higher than dipole.

In the following we shall comment on effects of crossing bonds 
on stability of the N\'{e}el ordered phase and the ferroquadrupole (spin nematic) phase. 
As was illustrated in FIG. 3-(a), 
there exist the N\'{e}el ordered phase and the ferroquadrupole phase 
which are separated by the line at $J_{1}(1)= 0$. 
By adding antiferromagnetic crossing bonds to the square lattice, 
it is clear that the N\'{e}el order exhibiting anti-alignment of spin is not stable. 
On the other hand, the quadrupole order is not alignment of spin but nematicity of spin, 
and from equation (\ref{U}) it can be seen that $O_{k,0}$ with even $k$ is even parity with respect to time reversal. 
The ferroquadrupole order is uniformly aligned nematic and does not seem to be suffer from geometrical frustration. 
So stability of  the ferroquadrupole phase is not affected 
by frustration due to antiferromagnetic crossing bonds unlike the N\'{e}el ordered phase. 
Now we set next nearest neighbor interactions $-J_{1}(\sqrt{2})=-\alpha J_{1}(1)$ and
$-J_{2}(\sqrt{2})=-\alpha J_{2}(1)$ with $0\le\alpha\le1/2$, 
then FIG. 3-(a) is expected to be changed as follows.
Phase boundary $J_{2}(1)=J_{1}(1)$ is unchanged by adding crossing bonds 
with ferromagnetic and ferroquadrupole interactions 
since the region $0\le J_{2}(1)< J_{1}(1)$ is saturated ferromagnetic ground state.
On the other hand, phase boundary $J_{1}(1)=0$ closes to the antiferromagnetic Heisenberg point
as $\alpha$ approaches $1/2$, i.e.,
by adding crossing bonds with antiferromagnetic and ferroquadrupole interactions, 
the ferroquadrupole phase becomes dominant
and the N\'{e}el ordered phase is suppressed 
if there do not exist different phases between these two phases. 
As for stability of  the N\'{e}el order, 
the triangular and pyrochlore lattices are slightly different situation 
from the square lattice with antiferromagnetic crossing bonds. 
The ground state phase diagrams of $S=1$ bilinear-biquadratic exchange model on the triangular 
and pyrochlore lattices are obtained in reference \cite{QT1,QT2,spinel2} 
and the same situation in the above scenario is shown. 

Finally we shall propose the physical realization of the ferroquadrupole phase 
in magnetic materials. 
Usually biquadratic interaction is small as compared with bilinear interaction and 
the ferroquadrupole phase is unphysical in magnetic materials. 
In reference \cite{mila} Mila and Zhang proposed a mechanism 
leading to a significant biquadratic interaction in $S=1$ systems as follows.
The super exchange interaction between atoms  
with three orbitals and two outer electrons per atom,  
which consists of the two singly occupied doubly degenerate orbitals with the lowest energy
and an unoccupied orbital with slightly higher energy.
The virtual electron transition via the higher energy orbital favors ferromagnetic spin interaction, which compensates largely the antiferromagnetic superexchange interaction. 
As a result, the biquadratic interaction becomes predominant relatively.  
Thus we expect that highly frustrated antiferromagnetic materials with biquadratic exchange interactions originated from the Mila-Zhang mechanism may exhibit the ferroquadrupole phase. 
 
\begin{acknowledgments}
I would like to thank Hosho Katsura and Akinori Tanaka for useful comments. 
\end{acknowledgments}

%
%

\nocite{*}


\begin{thebibliography}{}
\bibitem{aklt}I. Affleck, T. Kennedy, E. H. Lieb and H. Tasaki, Commun. Math. Phys. {\bf 155}, 477 (1988).
\bibitem{millet} P. Millet, F. Mila, F. C. Zhang, M. Mambrini, A. B. Van Oosten, V. A. Pashchenko, 
A. Sulpice, and A. Stepanov Phys. Rev. Lett. {\bf 83}, 4176 (1999).
\bibitem{mila}F. Mila and Fu-Chun Zhang, Eur. Phys. J. B {\bf 16} 7 (2000).
\bibitem{tsunetsugu}H. Tsunetsugu and M. Arikawa, J. Phys. Soc. Jpn. {\bf 75}, 083701 (2006).
\bibitem{QT1}A. L\"{a}uchli, F. Mila and K. Penc, Phy. Rev. Lett. {\bf 97} 087205 (2006).
\bibitem{QT2}S. Bhattacharjee, V. B. Shenoy and T. Senthil, Phys. Rev. B {\bf 74} 092406 (2006).
\bibitem{Yip}S. K. Yip, Phys. Rev. Lett.  {\bf 90} 250402 (2003).
\bibitem{imambekov}A. Imambekov, M. Lukin and E. Demler, Phys. Rev. A {\bf 68} 063602 (2003).
\bibitem{flavor}T. A. T\'oth, A. M. L\"{a}uchli, F. Mila and K. Penc, Phys. Rev. Lett. {\bf 105} 265301 (2010), Phys. Rev. Lett. {\bf 108} 029902 (2012).
\bibitem{pnic1}A. L. Wysock, K. D. Belashchenko and V. P. Antropov, Nat. Phys. {\bf 7} 485 (2011).
\bibitem{pnic2}R. Yu, Z. Wang, P. Goswami, A. H. Nevidomskyy, Q. Si and E. Abrahams, Phys. Rev. B {\bf 86} 085148 (2012).
\bibitem{HKT}K. Harada, N. Kawashima and M. Troyer, J. Phys. Soc. Jpn. {\bf 76} 013703 (2007).
\bibitem{TS}Tarun Grover and T. Senthil, Phy. Rev. Lett. {\bf 98} 247202 (2007).
\bibitem{nishiyama}Y. Nishiyama, Phys. Rev. B {\bf 83} 054417 (2011).
\bibitem{spinel1}K. Penc, N. Shannon, and H. Shiba, Phys. Rev. Lett. {\bf 93} 197203 (2004).
\bibitem{spinel2}E. Takata, T. Momoi, M. Oshikawa,  arXiv:1510.02373 (2015).
\bibitem{tu}H.-H. Tu , G.-M. Zhang and L. Yu Phy. Rev. B {\bf 74} 174404 (2006).
\bibitem{eckert}K. Eckert, \L. Zawitkowski , M. J. Leskinen, A. Sanpera and M. Lewenstein, N. J. Phys. {\bf 9} 133 (2007).
\bibitem{FKKI}Yu A. Fridman, O. A. Kosmachev, A. K. Kolezhuk and B. A. Ivanov, 2011 Phy. Rev. Lett. {\bf 106} 097202 (2011).
\bibitem{wei}T.-C. Wei, I. Affleck and R. Raussendorf, Phy. Rev. Lett. {\bf 106} 070501 (2011).
\bibitem{miyake}A. Miyake, Ann. Phys. {\bf 326} 1656 (2011).
\bibitem{LM}E. Lieb and D. Mattis, J. Math. Phys. (N. Y.) {\bf 3} 749 (1962).
\bibitem{munro}R. G. Munro, Phys. Rev. B {\bf 13} 4875 (1976).
\bibitem{parkinson}J. B. Parkinoson,  J. Phys. C:Solid State Phys. {\bf 10} 1735 (1977).
\bibitem{tanaka}A. Tanaka and T. Idogaki, Phys. Rev. B {\bf 56} 10774 (1997).
\bibitem{RQS1}X.-S. Ma, B, Daki\'c, W. Naylor, A. Zeilinger and P. Walther, Nat. Phys. {\bf 7} 399 (2011).
\bibitem{RQS2}X.-S. Ma, B. Daki\'c, S. Kropatschek, W. Naylor, Y.-H. Chan, Z.-X. Gong, L.-M. Duan, 
A. Zeilinger and P. Walther Nat. Phys. {\bf 4} 3583 (2014).
\bibitem{lieb-schupp1}E. H. Lieb and P. Schupp, Phys. Rev. Lett.  {\bf 83} 5362 (1999).
\bibitem{lieb-schupp2}E. H. Lieb and P. Schupp, Physica A {\bf 279} 978 (2000).
\bibitem{schupp}P. Schupp, arXiv:math-ph/0206021(2001).
\bibitem{LD}P. A. Ling\r{a}rd and O. Danielsen, J. Phys. C:Solid State Phys. {\bf 7} 1523 (1974).
\bibitem{FMH}W.-D. Fereitag and E. M\"{u}ller-Hartmann, Z. Phys. B Condensed Matter {\bf 88} 279 (1992).
\bibitem{depleted}M. Troyer, H. Kontani and K. Ueda, Phys. Rev. Lett. {\bf 76} 3822 (1996).
\bibitem{KLS}T. Kennedy, E. H. Lieb and S. Shastry, J. Stat. Phys. {\bf  53} 1019 (1988).
\bibitem{lieb}E. H. Lieb,  Phys. Rev. Lett. {\bf  62} 1201 (1989).
\bibitem{TTI}K. Tanaka, A. Tanaka and T. Idogaki, J. Phys. A: Mathe. and Gen. {\bf 34} 8767 (2001).
\bibitem{bc}When we consider global Hamiltonian constructed 
with translated copies of local Hamiltonian on bond sharing even numbered rings 
and open boundary condition in the direction parallel to the symmetry plane,  
bonds at the edges of the lattice in that direction are half weight, 
which should be replaced by full weight bonds. 
We can easily see that our conclusions are not affected by this modification 
since a collection of half weight bonds at the edges is also reflection symmetry .
\bibitem{KK}K. Kishi and K. Kubo, J. Phys. Soc. Jpn. {\bf 58} 2547 (1989).
\bibitem{woj}J. Wojtkiewicz, Eur. Phys. J. B {\bf 44} 501 (2005).
\bibitem{DLS}F. J. Dyson, E. H. Lieb and B. Simon, J. Stat. Phys. {\bf 18} 335 (1978).
\bibitem{NF}E. Jord\~{a}o Neves and J. Fernando Perez, Phys. Lett. A {\bf 114} 331 (1986).
\bibitem{QS1}H. H. Chen and P. M. Levy, Phys. Rev. Lett. {\bf 27} 1383 (1971), Phys. Rev. B {\bf 7} 4284 (1973).
\bibitem{QS2}V. M. Matveev, Sov. Phys. JETP {\bf 38} 813 (1974).
\bibitem{QS3}N. Papanicolaou, Nucl. Phys. B {\bf 240} 281 (1984).
\bibitem{QS4}K. Harada and N. Kawashima, Phys. Rev. B {\bf 65} 052403 (2002).
\bibitem{U}D. Ueltschi, J. Math. Phys. {\bf 54} 083301 (2013).
\bibitem{Lee}B. Lees, J. Math. Phys. {\bf 55} 093303 (2014).
\bibitem{note1}Here we consider Hamiltonian $U^{\dagger}{\cal H}(b)U$ 
under site-dependent filed $b(x)$ as equation (\ref{hb}), 
then we also see inequality for its ground state energy: $E(b)\ge E(0)$ similar 
to inequality (\ref{energy}). 
Setting $b(x)=b$ if $x\in\Lambda_{\rm L}$, otherwise $b(x)=-b$,
we also reach equation (\ref{ghb1}) similar to equation (\ref{icerule}). 
\bibitem{J1-J2}P. H. Y. Li, R. F. Bishop, and C. E. Campbell J. Phys.: Conf. Ser. {\bf 529} 012008 (2014).
\bibitem{honeycomb1}A. F. Albuquerque, D. Schwandt, B. Het\'{e}nyi, S. Capponi, 
M. Mambrini and A. M. L\"{a}uchli Phys. Rev. B  {\bf 84} 022406 (2011).
\bibitem{honeycomb2}P. H .Y. Li, R. F. Bishop, D. J. J. Farnell and C. E. Campbell Phys. Rev. B {\bf 86} 144404 (2012).
\end{thebibliography}

\end{document}